\documentclass[article]{revtex4}

\newcommand{\Riesz}{\left(-\hbar^2\triangle\right)^{\alpha/2}}
\newcommand{\eqref}[1]{(\ref{#1})}
\newcommand{\Ai}{\mathop{\mathrm{Ai}}}

\begin{document}

\title{On the nonlocality of the fractional Schr\"{o}dinger
equation}

\author{M. Jeng$^1$\footnote{mjeng@physics.syr.edu}, S.-L.-Y. 
 Xu$^1$\footnote{sxu01@physics.syr.edu}, E. Hawkins$^2$\footnote{mrmuon@mac.com} and 
J. M. Schwarz$^1$\footnote{jschwarz@physics.syr.edu}}

\affiliation{
$^1$Physics Department,
Syracuse University,
Syracuse, NY, 13244, USA
$^2$Department of Mathematics, University of York, UK}

\begin{abstract}
A number of papers over the past eight years
have claimed to solve
the fractional Schr\"{o}dinger equation for systems ranging from  
the one-dimensional infinite square well to the 
Coulomb potential to one-dimensional scattering with a rectangular barrier. 
However, some of the claimed solutions ignore the fact that the
fractional diffusion
operator is inherently nonlocal, preventing the fractional
Schr\"{o}dinger equation from being solved in the usual 
piecewise fashion. We focus on the one-dimensional infinite
square well and show that the purported groundstate, which
is based on a piecewise approach, is definitely not a
solution of the fractional Schr\"{o}dinger equation for
general fractional parameters $\alpha$.
On a more positive note, we present a solution to the fractional
Schr\"{o}dinger equation
for the one-dimensional 
harmonic oscillator with $\alpha=1$.

\end{abstract}

\maketitle


\section{Introduction}
A wide variety of stochastic processes are more general than the familiar Brownian motion, but presumably can still be described by modifying the diffusion equation using a fractional Laplacian operator~\cite{LevyIntro1,LevyIntro2}.
Such  ``fractional diffusion'' is now a
large and active field, and a number of books have been written
on the mathematics and physics of fractional
diffusion
operators~\cite{FractionalDiffusion1,FractionalDiffusion2,FractionalDiffusion3}.
In 2000, Laskin introduced
the fractional Schr\"{o}dinger equation, in which the
normal Schr\"{o}dinger equation is modified in analogy with
fractional diffusion~\cite{Laskin1,Laskin2,Laskin3}.
Laskin claimed to exactly solve this equation
in the case of the one-dimensional infinite square well~\cite{Laskin1}. A more recent (2006) work claimed to find solutions
again for the infinite one-dimensional square well (agreeing with Laskin's 
original solution), and
for one-dimensional scattering off of a barrier potential~\cite{Guo.Xu}.
A 2007 work used a different method of analysis to claim
solutions for the linear, delta function, and Coulomb
potentials in one dimension~\cite{Dong.Xu}.
 Laskin also recently built on
the same claimed solution to derive properties of the quantum
kernel~\cite{Laskin2005}.
The purpose of this work is to point out that of the many 
purported exact solutions presented in the literature, 
only the one for the delta function potential is correct.

The one-dimensional fractional Schr\"{o}dinger equation \cite{Laskin1} is
\begin{equation}
i\hbar \frac{\partial\psi (x,t)}{\partial t} =
D_\alpha \Riesz \psi (x,t)+
V(x,t)\, \psi (x,t),
\end{equation}
where $D_\alpha$ is a constant, 
$\Delta\equiv \partial^2/\partial x^2$ is the Laplacian, 
and $\Riesz$ is the quantum Riesz
fractional derivative:
\begin{equation}
\Riesz \psi(x,t) \equiv
\frac{1}{2\pi\hbar} \int_{-\infty}^{+\infty} dp\ 
e^{ipx/\hbar} \left| p \right|^\alpha \phi(p,t).
\end{equation}
Here, $\phi(p,t)$ is the Fourier transform of the
wavefunction, 
\begin{equation}
\phi(p,t) = \int_{-\infty}^{+\infty} dx\ \psi(x,t)\,
e^{-ipx/\hbar}.
\end{equation}
When $\alpha=2$, the quantum Riesz fractional
derivative becomes equivalent to an ordinary Laplacian, and
we recover the ordinary Schr\"{o}dinger equation.

We focus on the case where the potential is independent of
time, so we are interested in solutions of the following equation:
\begin{equation}
D_\alpha \Riesz \psi(x) + V(x)\, \psi(x) = E \psi(x).
\label{eq:time.indep}
\end{equation}
The fractional diffusion operator is a
{\it nonlocal} operator except when $\alpha=0,2,4,\ldots$.
This means that $\Riesz \psi(x)$ depends not just on
$\psi(y)$ for $y$ near $x$, but on $\psi(y)$ for all $y$.
This nonlocality, in turn, means that when solving Eq.~\eqref{eq:time.indep}, the form of the wavefunction in a
given region depends not just on the potential in that
region, but on the potential everywhere. Because of this,
for a piecewise-defined potential, we cannot follow the
normal strategy of solving separately for the wavefunction
in each piecewise region, and then using
conditions of continuity and differentiability to match up
the solutions. However, this is precisely
the strategy used in the papers cited above, and
the solutions obtained in those papers are thus invalid. We
illustrate the problem by looking in some detail at the case
of the one-dimensional infinite square well in Section II. The problems with the
purported solutions for other potentials are similar,
and are discussed in Section III. Section IV presents an exact solution for the one-dimensional fractional harmonic oscillator with $\alpha=1$, followed by conclusions in Section V. 


\section{Infinite one-dimensional square well}
\label{sec:square.well}
Consider Eq.~\eqref{eq:time.indep} in the limit of the potential becoming an infinite square well
\begin{equation}
V(x) = \left\{ 
\begin{array}{cll}
0 & \mathrm{if} & \left| x \right| < a \\
\infty & \mathrm{if} & \left| x \right| \ge a.
\end{array}
\right.
\end{equation}
We first note that for the case of free space, where the
potential $V$ is zero {\it everywhere}, it is easy to see
that plane waves are eigenfunctions of the
quantum fractional Hamiltonian:
\begin{equation}
\Riesz e^{ipx/\hbar} = \left| p \right|^\alpha e^{ipx/\hbar}.
\label{eq:plane.wave}
\end{equation}
However, Eq.~\eqref{eq:plane.wave}
is only valid if the function operated on is $e^{ipx/\hbar}$
{\it everywhere}; it is not a local equation that can be applied
just in a restricted region.
Because the quantum Riesz fractional derivative is a
nonlocal operator, the wavefunction in the well
knows about the wavefunction and potential outside of the
well.
Previous works looking at the one-dimensional infinite square
well incorrectly applied Eq.~\eqref{eq:plane.wave}
only inside the well, and concluded that
the solution inside the
well would be a simple linear superposition of 
left- and right-moving plane waves of the same
energy~\cite{Laskin1,Guo.Xu}.

Although these papers use an invalid assumption, could their end results
be correct nevertheless? Both papers claim that the
solutions for the one-dimensional square well are the
same for the fractional case as for the standard 
non-fractional
case, only with modified energies. So, they obtain for the
ground state
\begin{equation}
\psi_0(x) = \left\{ 
\begin{array}{ll}
A \cos \left(\frac{\pi x}{2a}\right) & 
\mathrm{for}\ \left| x \right| \leq a \\
0 & \mathrm{otherwise}.
\end{array}
\right.
\label{eq:ground.state}
\end{equation}
The Fourier transform of this is
\begin{eqnarray}
\nonumber
\phi_0(p) & := 
& \int_{-\infty}^{+\infty} dx\ e^{-ipx/\hbar} \psi_0(x), \\
& \,= & - \frac{A\pi\hbar^2}{a} 
\frac{\cos \left(ap/\hbar\right)}{p^2-(\pi\hbar/2a)^2}.
\label{eq:fourier.transform}
\end{eqnarray}
 From $\phi_0(p)$ we can calculate the fractional Riesz derivative:
\begin{equation}
\Riesz \psi_0(x) = 
- \frac{2A}{\pi} \left(\frac{\pi\hbar}{2a}\right)^\alpha 
\int_0^\infty dp\ \frac{p^\alpha}{p^2-1}
\cos\left(\frac{1}{2}\pi p\right)
\cos\left( \frac{\pi p x}{2a}\right).
\label{eq:Riesz.ground}
\end{equation}
We see here how the nonlocality manifests itself
in the mathematics. If we only looked at the wavefunction
inside the square well, then $\psi_0(x)$ would appear to consist
of plane waves of just two wavevectors, $\pm \pi / (2a)$.
However, in reality, $\psi_0(x)$ is $0$ outside the well,
making it a wave packet, rather than just a combination of
two plane waves, and so it contains a continuous range of
wavevectors, as seen in Eq.~\eqref{eq:fourier.transform}.
The fractional Riesz derivative thus sees all
these wavevectors.

Now we shall show that $\psi_0(x)$ is not a solution of the
infinite square well via a proof by contradiction. First,
assume that $\psi_0(x)$ is a solution of the fractional
Schr\"odinger equation. Then the fractional Riesz derivative
$\Riesz \psi_0(x)$ must be proportional to $\psi_0(x)$ on the open interval,
$|x|<a$, where $V(x)=0$.  
Since, $\psi_0(x)$ is continuous and $\psi_0(a)=0$, this implies that the limit $x\to a^-$ of \eqref{eq:Riesz.ground} should also vanish.  
However, this condition is \emph{not} equivalent to \eqref{eq:Riesz.ground}
vanishing at $x=a$ because the Hamiltonian includes the potential and
so we cannot rely on continuity of $\psi_0(x)$ at $x=a$.

A Fourier transform such as \eqref{eq:Riesz.ground} is
continuous if the integral is absolutely
convergent.
The integrand in \eqref{eq:Riesz.ground} is bounded by $p^\alpha$ for small $p$ and by $(1+\epsilon)p^{\alpha-2}$ for large $p$. Therefore, \eqref{eq:Riesz.ground} is indeed a continuous function for all $x$ for $-1<\alpha<1$.
Thus, for $-1<\alpha<1$, we can take the limit $x\to a^{-}$
\eqref{eq:Riesz.ground} by setting $x=a$,
and if $\psi_0(x)$ is a solution, this should give
zero:

\begin{equation}
f(\alpha):= \int_0^\infty dp\ \frac{p^\alpha}{p^2-1}
\cos^2\left(\frac{1}{2}\pi p\right) = 0.
\end{equation}
Taking the derivative with respect to $\alpha$,
we see
\begin{equation}
\frac{df}{d\alpha} = 
\int_0^\infty dp\ \frac{p^\alpha \ln p}{p^2-1}
\cos^2\left(\frac{1}{2}\pi p\right).
\label{eq:integral.derivative}
\end{equation}
The integrand in Eq.~\eqref{eq:integral.derivative}
is everywhere positive, so
$df/d\alpha>0$, and we cannot have
$f(\alpha)=0$ for all $\alpha$. 
The ground state \eqref{eq:ground.state} claimed in
Refs.~\cite{Laskin1} and~\cite{Guo.Xu} thus cannot be an
solution of the fractional Schr\"{o}dinger equation for 
all $\alpha$. It can only be a solution once in the interval
$-1<\alpha<1$---namely when $\alpha=0$.

The above argument does not hold for $1\le|\alpha|$ and in fact for some values of $\alpha$, \eqref{eq:Riesz.ground} is not continuous at $x=a$.  However, a related argument to one presented above
shows that this $\psi_0$ cannot be a solution at least for $1<\alpha<2$.    
For $\alpha=2$, on the other hand, $\psi_0(x)$ actually \emph{is} a
solution. Indeed it is a solution whenever the fractional Riesz
derivative is an ordinary derivative---that is, for
$\alpha=0,2,4,\ldots$. 

It may seem counterintuitive that
Eq.~\eqref{eq:ground.state} is not the correct ground state.
The standard ($\alpha=2$) Schr\"odinger equation for an
infinite potential well is equivalent to the Schr\"odinger
equation on an interval with the Dirichlet boundary
conditions $\psi(-a)=\psi(a)=0$. By raising that Hamiltonian
to the power $\alpha/2$ we get a plausible fractional
Laplacian and Eq.~\eqref{eq:ground.state} is indeed a solution. However, this is \emph{not} the Riesz fractional derivative. In other words, the fractional Schr\"odinger equation for an infinite potential well is not equivalent to the fractional Schr\"odinger equation on an interval. 

At this point, we do not know
what the true solutions are for values of $\alpha$ other than 
$0,\ 2,\ 4,\dots$.
In Ref.~\cite{Well.Numerical}, Zoia \emph{et al.} find numerical
solutions for the ground state. The solutions depend on
$\alpha$ and differ from the simple sine wave solution
in Eq.~\eqref{eq:ground.state}.


\section{Other systems}
While we have only discussed the infinite one-dimensional
square well in detail, the comments here equally invalidate
the other claimed solutions of the fractional
Schr\"{o}dinger equation. For example, in 
Section II of Ref.~\cite{Dong.Xu}, the linear potential,
\begin{equation}
V(x) = \left\{ 
\begin{array}{cll}
Fx & \mathrm{if} & x\geq 0 \\
\infty & \mathrm{if} & x<0
\end{array}
\right.
\qquad ,
\label{eq:linear.potential}
\end{equation}

\noindent is studied. The
authors of Ref.~\cite{Dong.Xu} treat this equation in a
piecewise approach by solving the equation for the potential
$V(x)=Fx$ and applying a boundary condition at $x=0$. This
is invalid for the same reasons stated above
for the square well potential. 
Similar comments apply
to the analysis of the Coulomb potential in section IV of
that same paper. 

Our comments, however, do not invalidate
the analysis of the delta
function potential in Section III of Ref.~\cite{Dong.Xu},
which did not implement a piecewise approach, but instead
worked with the Fourier transform of the delta function
potential.
However, the authors of Ref.~\cite{Dong.Xu} fail to note
that the bound state for the delta function potential is
valid only for $\alpha\geq 1$. For $\alpha\leq 1$, there is no bound state
since the integral in Eq. 33 of Ref.~\cite{Dong.Xu} diverges. In more recent
work, Dong and Xu~\cite{Dong.Xu.2008} attempt to solve the same problem again,
using a piecewise approach. They then compare this to their initial correct solution and derive an incorrect identity for the H-function.
 


\section{The fractional harmonic oscillator}

Consider the fractional Schr\"{o}dinger equation with the potential

\begin{equation}
V(x)=\frac{1}{2} kx^2.
\end{equation}
Fourier transforming Eq.~\eqref{eq:time.indep} gives
\begin{equation}
\frac{1}{2} k \hbar^2 \frac{d^2 \phi}{d p^2} =
(D_\alpha |p|^{\alpha} - E )\phi (p).
\label{eq:sho.momentum}
\end{equation}
In momentum space, the equation maps to the ordinary Schr\"{o}dinger 
equation with a positive $\alpha$ power law potential and $k=1/m$. In other
words, the kinetic
and potential energies have reversed roles. 

\subsection{WKB approximation} Given the mapping to ordinary quantum mechanics, in the limit
$\hbar\rightarrow 0$, one can use the WKB approximation in momentum space to
approximate the energy eigenvalues, with $p$ replacing $x$. The quantization condition in momentum space is 

\begin{equation}
\int_{p_1}^{p_2}\lambda(p)dp=(n+\frac{1}{2})\pi \hbar, \qquad n\in \{0\} \cup
\mathcal{Z}^+ \qquad ,
\end{equation}

\noindent where
$\lambda(p)=\sqrt{\frac{2}{k}(E-D_{\alpha}|p|^{\alpha})}$,
and $p_1$ and $p_2$ are the classical turning points, i.e.
$p_{1,2}=\pm
(E/D_{\alpha})^{1/\alpha}$. The above condition leads to 
\begin{equation}
E_n=\biggl(\frac{(n+\frac{1}{2})\hbar \pi \sqrt{k}
 (D_\alpha)^{1/\alpha}\Gamma(\frac{3}{2}+\frac{1}{\alpha})}{2\sqrt{2}\Gamma(\frac{3}{2})\Gamma(1+\frac{1}{\alpha})}\biggr)^{\frac{2\alpha}{2+\alpha}}.
\end{equation}
This agrees with Laskin's more general WKB result~\cite{Laskin2002} for an arbitrary power-law potential. However, by the argument we have just given, this special case is better justified than Laskin's general claim.
It is unclear whether there is any reason to believe the WKB
approximation to be vaild for systems other than the
harmonic oscillator, since for other potentials, the
Fourier transform of the fractional 
Schr\"{o}dinger equation will not be an ordinary
Schr\"{o}dinger equation.

\subsection{An exact solution for $\alpha=1$}
When $\alpha=1$, Eq.~\eqref{eq:sho.momentum} becomes 
\begin{equation}
\frac{1}{2} k \hbar^2 \frac{d^2 \phi}{d p^2} =
(D_1 |p| - E )\phi (p) \ .
\label{eq:sho.Cauchy.momentum}
\end{equation}
Restricting to $p>0$ or $p<0$, this differential equation is equivalent to the Airy equation
(by a rescaling transformation).
For a normalizable wavefunction,
we must have
$\phi(p=\infty)=\phi(p=-\infty)=0$.
This condition rules out Airy functions of the second
kind ($\mathrm{Bi}(z)$) as solutions. Because of the symmetry of
the potential, the solutions are alternately symmetric or antisymmetric. More precisely, 
\begin{equation}
\phi (p) = (\mathop{\mathrm{sgn}} p)^n \Ai(\kappa |p| - r_n) \ ,
\end{equation}
where 
$\kappa\equiv \left( 2 D_1 / (k \hbar^2) \right)^{1/3}$, and
the $r_n$'s are the successive roots of $\Ai'$ (for $n$
even), or of $\Ai$ (for $n$ odd). 
The energy eigenvalues are
\begin{equation}
E_n = - \left( \frac{k}{2} \hbar^2 D_1^2 \right)^{1/3} r_n .
\label{eq:show.Cauchy.energies}
\end{equation}
Note that $r_n<0$, so $E_n>0$.

Using the asymptotic expansion of the Airy function, we find
that the roots are well approximated by
\begin{equation}
r_n \approx - \left( \frac{3\pi}{4} \left( n + \frac{1}{2}
\right) \right)^{2/3}, \qquad n\in \{0\} \cup
\mathcal{Z}^+
\label{eq:roots.Airy.derivative}
\end{equation}
in the limit of large $n$. 
This approximation reproduces the result of the
WKB approximation above. 
This approximate formula
is off by 8.7\% for $n=0$, 0.77\% for $n=1$, 
and 0.41\% for $n=2$, and rapidly becomes more accurate for
larger $n$. Inserting Eq.~\eqref{eq:roots.Airy.derivative}
into Eq.~\eqref{eq:show.Cauchy.energies} thus gives a very good
approximate formula for the energies of the simple harmonic
oscillator for $\alpha=1$.

\section{Conclusions}
It would be useful to know
the correct groundstate
of the one-dimensional infinite square well 
or harmonic oscillator for
general $\alpha$, but this is a difficult problem.
In Ref.~\cite{Well.concavity.proof},
Ba\~{n}uelos \emph{et al.} needed a lengthy proof merely to show
that the groundstate solution for the 
infinite square well in the region $(-1,1)$ is concave on
the interval
$(-\frac{1}{2},+\frac{1}{2})$.

Similar technical issues regarding nonlocality have arisen in the statistical mechanics
community as well. For example, in Ref.~\cite{levy1}, the
authors claimed to analytically determine the mean first
passage time for a L\'evy flight with absorbing boundary
conditions on the interval $[0,1]$. They did so by imposing
the standard absorbing boundary conditions for the
probability density at $x=0$ and $x=L$. However, a
subsequent publication \cite{levy2} pointed out that
due to the nonlocal nature of the L\'evy flight, the
correct boundary condition is for the probability density to
vanish for all $x\le 0$, and for all $x\ge L$, rendering the
analysis in Ref.~\cite{levy1} invalid. 

Finally, one must also ask about possible physical 
realizations of the fractional Schr\"odinger equation.  
In Ref. \cite{Well.Numerical}, Zoia, Rosso, and Kardar constructed a lattice model whose continuum limit is described by fractional diffusion using symmetric Toeplitz matrices. A quantum representation of this model via a mesoscopic network of long range connections whose hopping amplitudes are described by the entries of these matrices may be a realization of the fractional Schr\"odinger equation.
Further modification could lead to experimental tests of the problems we have discussed here.

JMS would like to thank the hospitality of the Aspen Center for Physics where part of this work was completed and support from NFS-DMR-0645373.

\end{document}